\RequirePackage{snapshot}
\documentclass[Afour,times,sageh,doublespace]{sagej}

\usepackage{graphicx}
\usepackage{subfig}
\usepackage{hyperref}
\usepackage{booktabs}
\usepackage{siunitx}
\usepackage{enumitem}
\usepackage{color}

\definecolor{textR1}{rgb}{0.0,0.0,0.0}
\definecolor{textR2}{rgb}{0.0,0.0,0.0}
\definecolor{textR3}{rgb}{0.0,0.0,0.0}

\title{Studies on the energy and deep memory behaviour of a cache-oblivious, task-based hyperbolic PDE solver}

\runninghead{Charrier et al.}

\author{Dominic E.~Charrier\affilnum{1}, 
  Benjamin Hazelwood\affilnum{1}, 
  Ekaterina Tutlyaeva\affilnum{2},
  Michael Bader\affilnum{3},
  Michael Dumbser\affilnum{4},
  Andrey Kudryavtsev\affilnum{5}, 
  Alexander Moskovsky\affilnum{2},
  Tobias Weinzierl\affilnum{1}
}

\affiliation{\affilnum{1} Durham University, Department of Computer Science, Stockton
  Road, DH13LE Durham, Great Britain\\
  \affilnum{2} RSC Group, Kutuzovskiy av.~36, 121170 Moscow, Russia\\
  \affilnum{3} Technische Universit\"at M\"unchen, Department of Informatics,
  Boltzmannstr.~3, 85748 Garching, Germany\\
  \affilnum{4} University of Trento, Dipartimento di Ingegneria Civile
  Ambientale e Meccanica, Via Mesiano~77, 38123 Trento, Italy\\
  \affilnum{5} Intel, 1900 Prairie City Rd, 95630 Folsom, United States
}

\corrauth{Tobias Weinzierl, Durham University, Department of Computer Science, Stockton Road, Durham, DH1 3LE, Great Britain}
\email{tobias.weinzierl@durham.ac.uk}

\newcounter{observationcounter} 
\newenvironment{observation}[1][Observation]
{\vspace{0.2cm}\noindent\textbf{#1 \arabic{observationcounter}. }\
\stepcounter{observationcounter} }{\vspace{0.2cm}}

\begin{document}

\begin{abstract}
 We study
\textcolor{textR2}{
 the performance behaviour of 
}
a seismic simulation using the ExaHyPE engine
\textcolor{textR2}{
 with a specific focus on memory characteristics and energy needs. 
}
ExaHyPE combines dynamically adaptive mesh refinement (AMR) with ADER-DG.
It is parallelized using tasks, and it is cache efficient.
AMR plus ADER-DG yields a task graph which is highly dynamic in
nature and comprises both arithmetically expensive 
\textcolor{textR2}{
 tasks and tasks which challenge the memory's latency.
}
The expensive tasks and thus the whole
code benefit from AVX vectorization,
though we suffer from memory access bursts.
A frequency reduction of the chip improves the code's energy-to-solution.
Yet, it does not mitigate burst effects.
The bursts' latency penalty becomes worse once we add
%
\textcolor{textR3}{
 Intel\textregistered\ Optane\texttrademark\ technology,
 increase the core count significantly, or
 make individual, computationally heavy tasks fall out of close caches.
}
Thread overbooking to hide away these latency penalties 
\textcolor{textR1}{becomes}
 contra-productive
\textcolor{textR1}{
 with non-inclusive caches as it 
}
destroys the cache and vectorization character.
In cases where memory-intense and computationally expensive tasks
overlap, 
ExaHyPE's cache-oblivious implementation 
\textcolor{textR1}{nevertheless}
can exploit 
\textcolor{textR3}{
 deep, non-inclusive, heterogeneous memory
}
\textcolor{textR2}{
effectively, as main memory misses arise infrequently
and slow down only few cores.
}
We thus propose that upcoming supercomputing simulation codes with dynamic,
inhomogeneous task graphs are actively supported by thread runtimes 
\textcolor{textR1}{
 in intermixing tasks of different compute character, and we propose that future 
 hardware actively allows codes to downclock the cores running particular task
 types.
}

\end{abstract}

\maketitle 

\section{Introduction}

\begin{figure}[htb]
 \begin{center} 
  \includegraphics[width=0.32\textwidth]{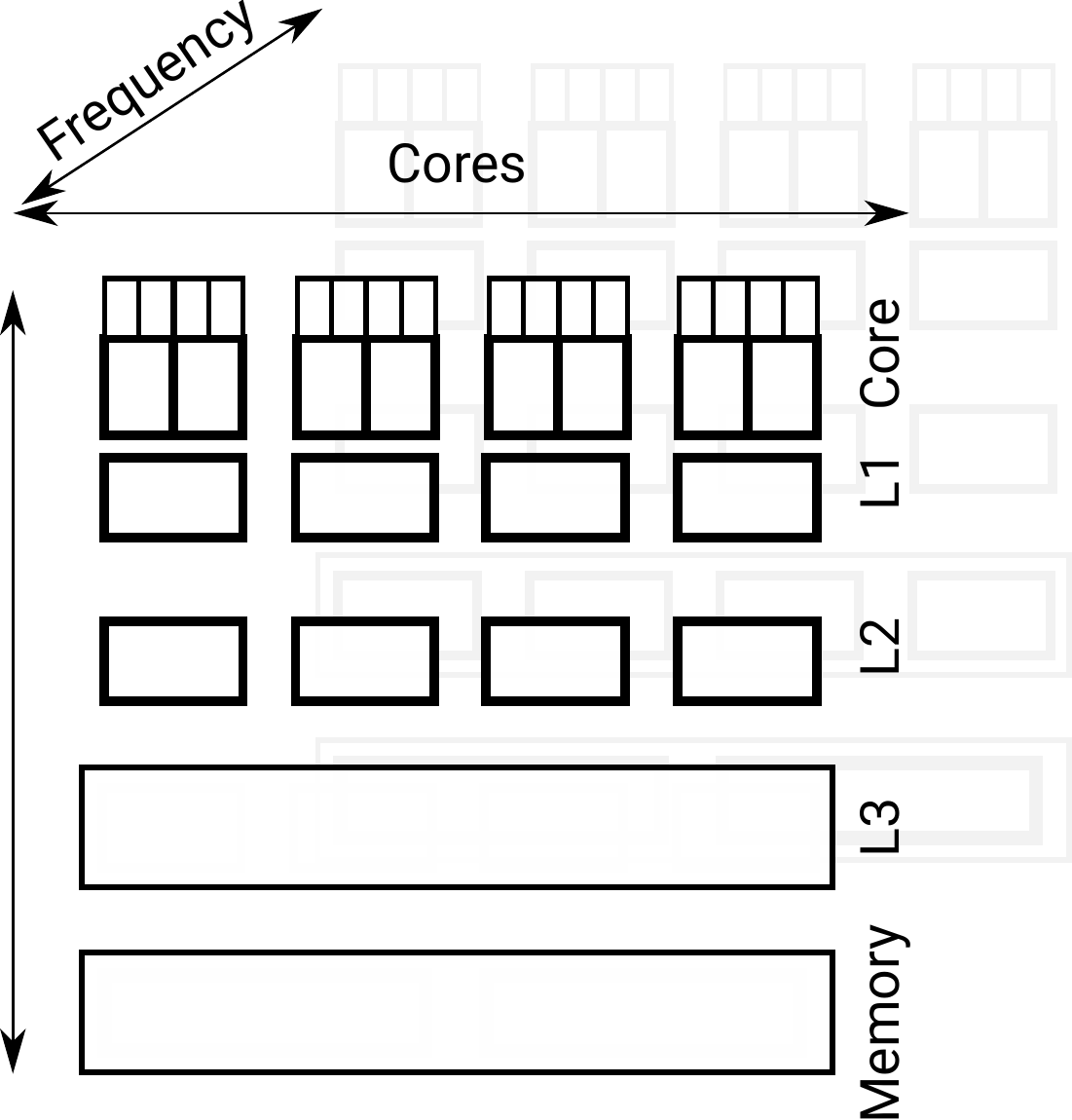}
 \end{center}
 \vspace{-0.2cm}
 \caption{
  \textcolor{textR2}{
   Vertical inhomogeneity of the data access cost, and thus speed, arises from
   multiple cache levels and different cache coherence strategies (inclusive
   vs.~non-inclusive).
   With \textcolor{textR3}{the Intel\textregistered\ Optane\texttrademark\
   technology}, main memory
   \textcolor{textR3}{effectively}
   becomes a fourth cache and an additional memory layer is added at the
   bottom.
   Horizontal inhomogeneity arises from the fact that memory is logically shared
   yet physically distributed.
   Further diversity stems from the fact that a core hosts multiple
   (hyper-)threads which in turn might accommodate multiple logical threads.
   A third diversity dimension is introduced by the opportunity to calibrate the
   cores' frequency.
  }
  \label{figure:introduction:diversity}
  \vspace{-0.5cm}
 }
\end{figure}

%
%
The memory architectures in mainstream supercomputing (Intel-inspired
architectures) become more and more inhomogeneous.
We classify these trends into clocktick, vertical and horizontal inhomogeneity
\textcolor{textR2}{(Figure~\ref{figure:introduction:diversity})}.
\textcolor{textR1}{
Modern architectures can modify the chip frequencies of some components.
}
Modern memory hierarchies are built in layers with the chip's registers on
the top, persistent memory at the bottom, \textcolor{textR1}{and caches
in-between}.
This yields \textcolor{textR2}{the} vertical dimension.
As access from the CPU registers to the main memory is very expensive, the
caches hold data copies temporarily.
Small intermediate memory layers can deliver data quick to
the cores.
\textcolor{textR1}{
Modern chips are predominantly multi-socket systems.
This yields \textcolor{textR2}{the} horizontal dimension.
}
Though the main memory and some intermediate memory layers are logically
shared between all cores, the chip technically is split up into sets of
cores with their own memories and memory controllers.
Data access cost within one layer \textcolor{textR2}{depends} on whether
data resides on the local segment of memory or has to be fetched from memory
technically associated to other sets of cores.

%
%
Neither vertical and horizontal nor frequency diversity are new.
Their character however evolves and their impact on code
performance increases.
There are at least three \textcolor{textR1}{recent} hardware trends
to \textcolor{textR1}{consider}:
With an increase of core counts, \textcolor{textR1}{non-uniform memory access}
(NUMA) effects gain importance.
More cores and their caches have to be synchronised, while the pressure on the
main memory increases.
With the arrival of more inhomogeneous or new beyond-main memory storage
technology (Intel\textregistered\ Optane\texttrademark\ technology or MCDRAM)
\textcolor{textR1}{which introduce new cache layers,
as well as with the farewell of inclusive caching with the 
Intel\textregistered\ Xeon\textregistered\ Scalable processors (Skylake)---though likely to be compensated to some
degree with the advent of mesh interconnects---}we witness \textcolor{textR1}{increased} non-uniformity when it comes to memory accesses.
With the opportunity to down- or upclock system components---either
triggered by users or the energy controllers on board---we finally face
further fluctuations in effective speed.

\begin{figure}[htb]
 \begin{center} 
  \includegraphics[width=0.49\textwidth]{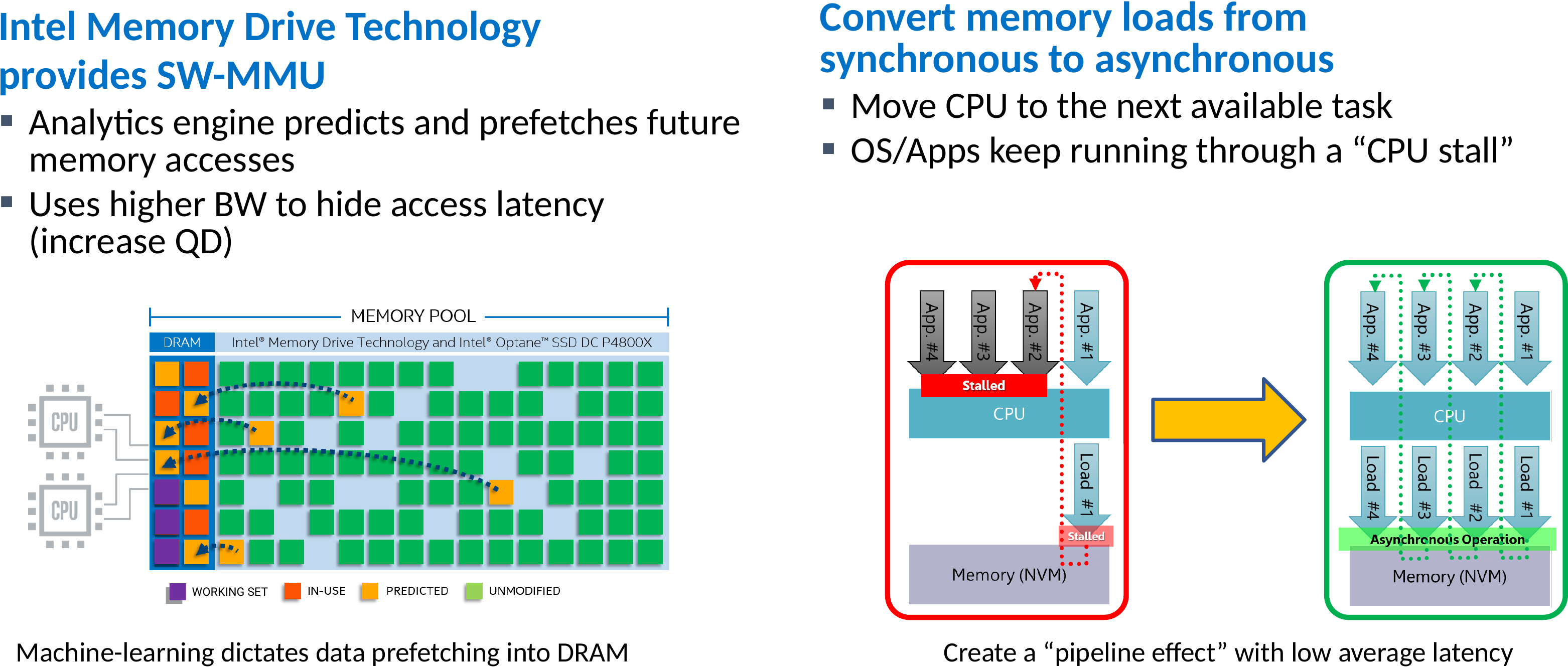}
 \end{center}
 \vspace{-0.2cm}
 \caption{
  \textcolor{textR2}{
   Intel\textregistered\ Optane\texttrademark\  SSD DC P4800X Series with Intel\textregistered\
   Memory Drive Technology operation mode. It relies on software IP for memory management. 
   Newer
   products such as Intel\textregistered\ Optane\texttrademark\  DC Persistent
   Memory
   are \textcolor{textR3}{promised to integrate}  
   on DIMM
   form-factor and also 
   \textcolor{textR3}{to} introduce 
   a memory mode fully managed by the CPU without extra
   software\textcolor{textR3}{; and hence smaller cost penalty}.
  }
  \label{figure:introduction:optane-imdt}
  \vspace{-0.5cm}
 }
\end{figure}

%
%
\textcolor{textR1}{These hardware features} are imposed on HPC simulations by the vendors.
\textcolor{textR2}{Despite co-design efforts,}
\emph{algorithmic and hardware
evolution seem to diverge for some of the most advanced simulation codes.}
\textcolor{textR1}{
Code developer invest significant development time into the vectorization of
their core compute kernels. 
Downclocking hits vectorization.
Code developers invest significant time into a flexible task decomposition of
their codes to uncover the maximum concurrency.
Yet, modern numerics yield task graphs that have heterogeneous compute
characteristics, non-predictable runtime cost and dependencies that change
frequently.
An example are predictor-corrector schemes with expensive predictors and cheap
correctors, built on top of Newton- or Picard-iterations with dynamic
termination criteria and dynamic adaptive mesh refinement
(AMR).
Horizontally diverse multicore systems challenge NUMA-aware scheduling.
Finally, developers invest significant time into cache blocking and
compute routines of high arithmetic intensity which
exploit all vector registers.
}
New memory layers typically deliver improved bandwidth
and storage size, but also increase latency.
For embarrassingly parallel codes streaming data through the cores as we find
them in machine learning\textcolor{textR1}{,} in-memory database systems
\citep{Boyandin:18:OptaneInMemoryDB}
\textcolor{textR1}{or dense matrix-matrix multiplications}
\citep{Kudryavtsev:17:NUMAOptane}, 
\textcolor{textR1}{
 latency poses a manageable challenge:
}
threads accessing remote memory are postponed and switched with other threads.
We ``asynchronize'' threads and memory accesses.
Intel's IMDT is explicitly built with this in mind
(Figure~\ref{figure:introduction:optane-imdt}).
\textcolor{textR1}{
 Indeed, moving data into the main memory upon request can even improve the
 performance, as moving the data into the ``right'' memory location eliminates
 NUMA penalties without complicated first-touch optimisations
}
\citep{Kudryavtsev:17:NUMAOptane}.
Such a programming model, being similar to CUDA, 
\textcolor{textR1}{
is however problematic for codes which are tailored towards cache
reusage, exploit all registers, and thus suffer from context switches.
}

\textcolor{textR1}{
 Our case study on a complex AMR code with a non-homogeneous task pattern
 showcases flavours of this hardware-software divergence.
 It suggests that memory latency becomes a major showstopper.
%
%
 Unfortunately, 
 (i) frequency modifications are ill-suited to
 tackle the latency problem---they help to improve the energy efficiency
 though---(ii) task/thread oversubscription is ill-suited to cure it either if
 data swapped out is not reliably backed up in the next-level cache, and (iii)
 additional memory layers amplify latency penalties.
 We however uncover that a heterogeneous task graph where tasks of
 different computational character are intermixed reduces the memory pressure
 and latency penalty.
 As a consequence, our code performs, from a memory point of view, better with 
 dynamic AMR than with regular grids once the task character difference
 (compute- vs.~memory-heavy) is reasonably high and dynamic AMR starts to
 mix those different tasks.
 This counter-intuitive result is, to the best of our knowledge, novel, and our
 report also seems to be the first in a line that studies the impact of the
 Intel Optane technology on a non-trivial solver for partial differential
 equations (PDEs) from both a performance and an (memory) energy consumption view.
}

%
%
The case study is structured as follows:
We give an overview over our benchmark code base ExaHyPE, before we 
\textcolor{textR1}{phrase our research hypotheses.
They circumscribe common expectations and will, in large parts, be falsified
by our results. 
Our text next
}
describes the two test machines.
In \textcolor{textR1}{
the subsequent section, we benchmark the code's runtime
characteristics, before we try to find evidence for our hypotheses.
The findings are summarized in our conclusion and guide future work.
}


\section{The ExaHyPE benchmark code}
\label{section:benchmark-code}

\begin{figure}[htb]
  \centering
  \includegraphics[width=0.5\textwidth]{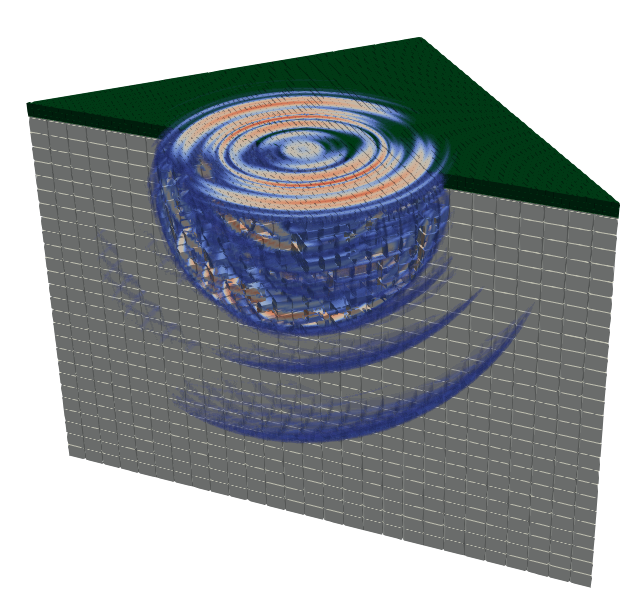}
  \caption{
   Cut through the solution of the LOH.1
   below the surface. Waves propagate from this point but
   benchmark. A point source induces an ``earthquake'' just
   yield complicated patterns as the cubic domain contains two layers of
   different material~\citep{SPICE:06:LOH1}.
   \textcolor{textR2}{
   Regular grid visualization though the experiments run with both regular and
   dynamically adaptive meshes.
   }
   \label{fig:loh1}}
\end{figure}

\textcolor{textR2}{Our experiments study a strongly simplified and idealized 
earthquake scenario (Figure~\ref{fig:loh1})}
realized through the ExaHyPE engine \citep{Software:ExaHyPE}.
ExaHyPE solves hyperbolic differential equations 
\textcolor{textR1}{
in their first order formulation
}
with ADER-DG.
ADER-DG is a predictor-corrector scheme
\citep{Charrier:18:AderDG,Dumbser:06:ADERDG} which traverses a grid tessellating the computational domain and first computes per mesh cell a
predicted solution evolution.
This prediction is of the same order in time as the spatial order (typically
$p \in \{3,4,\ldots,9\}$) and is determined implicitly.
Solving such an implicit space-time problem is computationally possible as we
neglect the solution in neighbouring cells.
It is a cell-local prediction.
The implicit solve of high order renders the predictor computationally intense.
Riemann solves in a second step tackle the arising jumps in the predicted
solutions along cell faces, before a corrector step sums up the result of the
prediction and the Riemann solves.
These two follow-up steps are arithmetically cheap.

ExaHyPE employs a dynamically adaptive Cartesian grid.
It is constructed from a spacetree \citep{Weinzierl:11:Peano,Weinzierl:18:Peano}.
The term spacetree describes a generalization of the octree/quadtree concept.
The meshing supports
dynamically adaptive grids which may change in each and every time step.
On purpose, we thus neglect multi-node runs.
They inevitably yield
load balancing challenges which hide the per-node memory effects
studied here.
The mesh topology determines which ADER-DG tasks can be run in parallel
\citep{Charrier:18:Enclaves}:
\textcolor{textR1}{All Riemann solves for example} are embarrassingly parallel
but require input from their two neighbouring cells. 
Along adaptivity boundaries, more than two cells are involved.
As the adaptivity changes, every time step induces a different multicore task
pattern, i.e.~the sequence and structure of parallel work items never is the
same between any two time steps.
We do not have a invariant task graph.

Combining a predictor-corrector scheme with task-based parallelism implies that 
(i) very compute-intense steps take turns with tasks that are computationally
cheap;
(ii) the memory demands change as dynamic mesh
refinement allocates additional blocks in the main memory while mesh coarsening 
releases memory segments;
(iii) the concurrency profile of the code is time-dependent and changing such
that dynamic tasking with task stealing is required.
To cope with these characteristics, our code base is subject to three
optimizations.

\subsection{Homogenization of the task execution}

In ADER-DG, all cell-based (correction and prediction) tasks are independent of
each other cells.
All Riemann tasks are independent of each other, too.
Between those types, dependencies exist: 
A predictor requires the input from the correction which in turn requires the
result of $2d$ Riemann solves. 
Each Riemann solve requires input from the two predictions of adjacent cells.

ExaHyPE offers two task processing modes.
In its basic variant, it first issues one type of task, processes these tasks
(which are all independent of each other), and then continues with the next type
of tasks.
\textcolor{textR1}{
Per time step, it first spawns all predictor tasks, then all Riemann tasks, and
finally all corrector tasks.
}
Dynamic adaptivity introduces additional 
\textcolor{textR1}{grid modification tasks}.
Each sweep is homogeneous w.r.t.~its compute profile.
The total time step however exhibits inhomogeneous character.
This scheme is equivalent to a breadth-first traversal
of the task graph.

An alternative variant is the fused mode \citep{Charrier:18:AderDG}.
It issues a task as soon as its input data are available.
A Riemann solve starts as soon as the predictions of the two
adjacent cells become available---it does not wait for all predictions to
terminate---and a corrector task is issued immediately once all $2d$
Riemann solves on the cell's adjacent faces are solved.
We issue tasks as soon as possible. 
\textcolor{textR1}{This induces}
some overhead to find
out whether a task is already ready.
\textcolor{textR1}{Yet, the} latter approach allows 
\textcolor{textR1}{
the task runtime to orchestrate tasks of different type to run concurrently.  
}
This homogenizes, i.e.~averages the character of the tasks over a time
step.
Though we have no absolute control on the processing order of the tasks---this
is up to the task runtime---we may assume that the task graph is processed
close to a depth-first order \citep{Reinders:07:TBB}.

\subsection{Temporal and spatial blocking}

ADER-DG's predictor \textcolor{textR1}{inherently} realizes spatial and temporal
blocking of data accesses \citep{Kowarschik:03:CacheTechniquesOverview}:
The expensive implicit solves are not \textcolor{textR2}{run} on the whole mesh
but on a per-cell basis.
This means many floating point operations are executed over a relatively small
set of data.
On top of this, ExaHyPE's grid traversal localizes all data accesses further.
It traverses the grid along a space-filling curve whose
H\"older continuity yields a spatial and temporal locality of data accesses
\citep{Weinzierl:18:Peano}.
If the result of a Riemann solve feeds into once adjacent cell, the probability
that the second adjacent cell is processed shortly after is high.
Together with the optimization resulting from the homogenization, ExaHyPE
realizes a cache-oblivious algorithm,
\textcolor{textR1}{
 which fuses correction and prediction.
 These cell operations are executed directly after another and merged into one
 task.
}

\subsection{Optimization of task core routines}

ExaHyPE \textcolor{textR1}{customizes} the engine:
As soon as architecture, number of equation unknowns, PDE term
types and polynomial orders are known---as they are for our LOH.1
setup---an ExaHyPE preprocessor (toolkit) can rewrite the
most time-consuming code parts into manually vectorized, tailored code kernels.
For these, it employs AVX instructions, appropriate alignment and padding,
as well as aggressive function inlining:
the application-generic engine machinery is rewritten without 
virtual function calls.

\section{Research hypotheses}

\textcolor{textR1}{
We consider our ExaHyPE benchmark to be a prime candidate to study and assess
new memory hierarchies, as the code exhibits multiple characteristic properties
of modern simulation software:
}
First, high-order, locally implicit approaches are one 
\textcolor{textR1}{popular}
way forward to exploit
vectorization.
\textcolor{textR1}{
Second, we expect many future simulation codes to consist of different task
types.
Computationally demanding tasks---the workhorses---take turns with other,
cheaper tasks which are however mandatory for advanced numerics.
We focus on a predictor-corrector scheme here. 
Another popular example for such an algorithmic imprint are
multigrid codes with expensive fine grid smoothers and cheap coarse equation
systems.
Third, we expect the majority of future codes to exploit some kind of dynamic
adaptivity to invest compute power where it pays off most.
As a result, task graphs, memory footprint and compute facility
needs are never invariant or temporarily homogeneous.
Notably, we assume that proper a priori prefetching becomes very difficult or
even impossible for the runtimes.
Forth, we assume that cache blocking---realised here
implicitly through a computationally heavy predictor which acts on one cell of
the mesh only---removal of virtual function calls, padding, manual vectorization
and so forth are state-of-the-art for any compute kernel.
}

By means of our non-trivial benchmark setup, we follow up on
\textcolor{textR1}{the following hypotheses:
\begin{enumerate}[noitemsep,leftmargin=0.3cm,topsep=0.2cm]
  \item A frequency increase of the compute units helps to improve the 
  time-of-solution, while a frequency reduction improves the energy-of-solution
  ratio. It also weakens the latency penalty.
  \item Task oversubscription helps to hide latency
  effects. The popular (light) oversubscription pattern from CUDA enters
  mainstream processors.
  \item The increased latency introduced by additional memory layers
  (fourth-level cache) harms notably those runs that exhibit a strongly
  inhomogeneous data access pattern (dynamic AMR).
  In-hardware prefetching breaks down.
\end{enumerate}
}

\section{Benchmark setup and system}
\label{section:benchmark-system}

Our benchmark \textcolor{textR1}{is}
\textcolor{textR3}{run on the following} 
server configurations. 
The first one is an Intel Xeon E5-2650V4 (Broadwell) cluster in a two socket configuration with 24 cores.
\textcolor{textR2}{They} run at 2.4~GHz.
TurboBoost increases this to up to 2.8~GHz, but a core executing AVX(2)
instructions might fall back to a minimum of 1.8~GHz to stay within the TDP
limits~\citep{Microway:18:Broadwell}. 
Each node has access to 64~GB of \textcolor{textR2}{2.4~MHz} TruDDR4 memory.

\textcolor{textR1}{
The second \textcolor{textR2}{configuration} is \textcolor{textR2}{a}
dual-socket Intel Xeon Scalable Gold 6150 with 18 physical cores per
socket, clocked at 2.70~GHz, and equipped with 192 GB of \textcolor{textR2}{2.7~MHz} DDR4 memory (12 ranks of
16~GB modules).
The chip may reduce the base clock frequency to 2.3~GHz for
AVX2 and to 1.9~GHz for AVX-512 \citep{Intel:18:XeonGold}.
The other way round, Thermal Velocity Boost permits individual cores to upclock
to up to 3.7~GHz temporarily.
}


%
%
\textcolor{textR3}{
In our experiments, the latter system is expanded with $6 \times$ 
Intel DC Optane SSD P4800X, i.e.~375~GB, non-volatile memory.
The SSDs are connected via PCIe-switch IC to the CPU, while
the Intel Memory Drive Technology 
implements software-defined memory (SDM) on-top of the Intel Optane
technology SSDs (cmp.~Fig.~\ref{figure:introduction:optane-imdt}). 
This Intel Memory Drive Technology (IMDT) uses part of the overall
memory capacity from the DRAM for caching, prefetching, and endurance protection, i.e.~the drives
become transparently available to the operating system as system memory.
Though our memory totals to roughly 1.4~TB, we stick to the default IMDT settings recommended by Intel which limits the available memory to $8\times$ the main memory.
Larger ratios than $1:8$ would lead to performance drops according to the vendor.
}

All shared memory parallelization relies on Intel's Threading Building
Blocks (TBB)~\citep{Reinders:07:TBB} while Intel's 2018 C++ compiler translated all codes.
We use Likwid \citep{Treibig:10:Likwid} to read out
hardware \textcolor{textR1}{and energy}
counters made available through RAPL.
\textcolor{textR3}{On the Purley platform chip,}
we use energy sensors \textcolor{textR2}{that} are directly
attached to the board.

\textcolor{textR2}{
Our experiments study the LOH.1 benchmark
\citep{Day:01:LOH1,SPICE:06:LOH1} realized through the ExaHyPE engine \citep{Software:ExaHyPE}.
LOH.1's artificial setup splits up 
a cubic domain into two horizontal layers of material.
An
earthquake then is induced as point source inside the cube.
Sensors close to the domain surface track incoming waves.
While LOH.1 is artificial, it exhibits real-world simulation
characteristics with its material transition, a source term and non-trivial
inference and reflection patterns.
\textcolor{textR1}{To obtain high quality results at reasonable cost,}
a feature-based refinement criterion follows the steepest solution gradients
and shocks.
The mesh spreads from the point source.
}

\section{Benchmark code characteristics}
\label{section:code-characteristics}

\subsection{Automatic frequency alterations}
\label{subsection:measurements:frequency}

\begin{figure*}[htb]
 \begin{center}
  \includegraphics[width=0.3\textwidth]{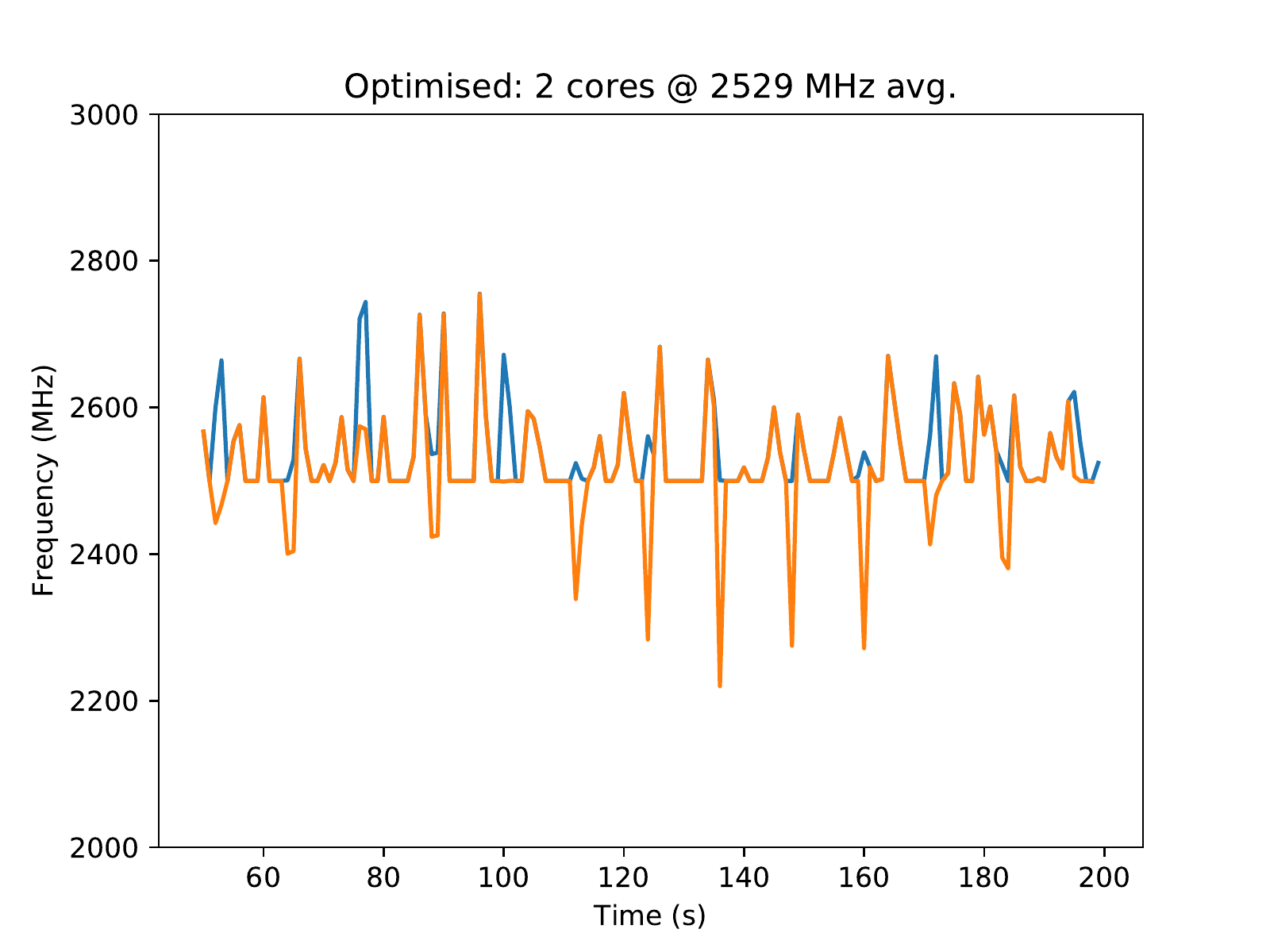}
  \includegraphics[width=0.3\textwidth]{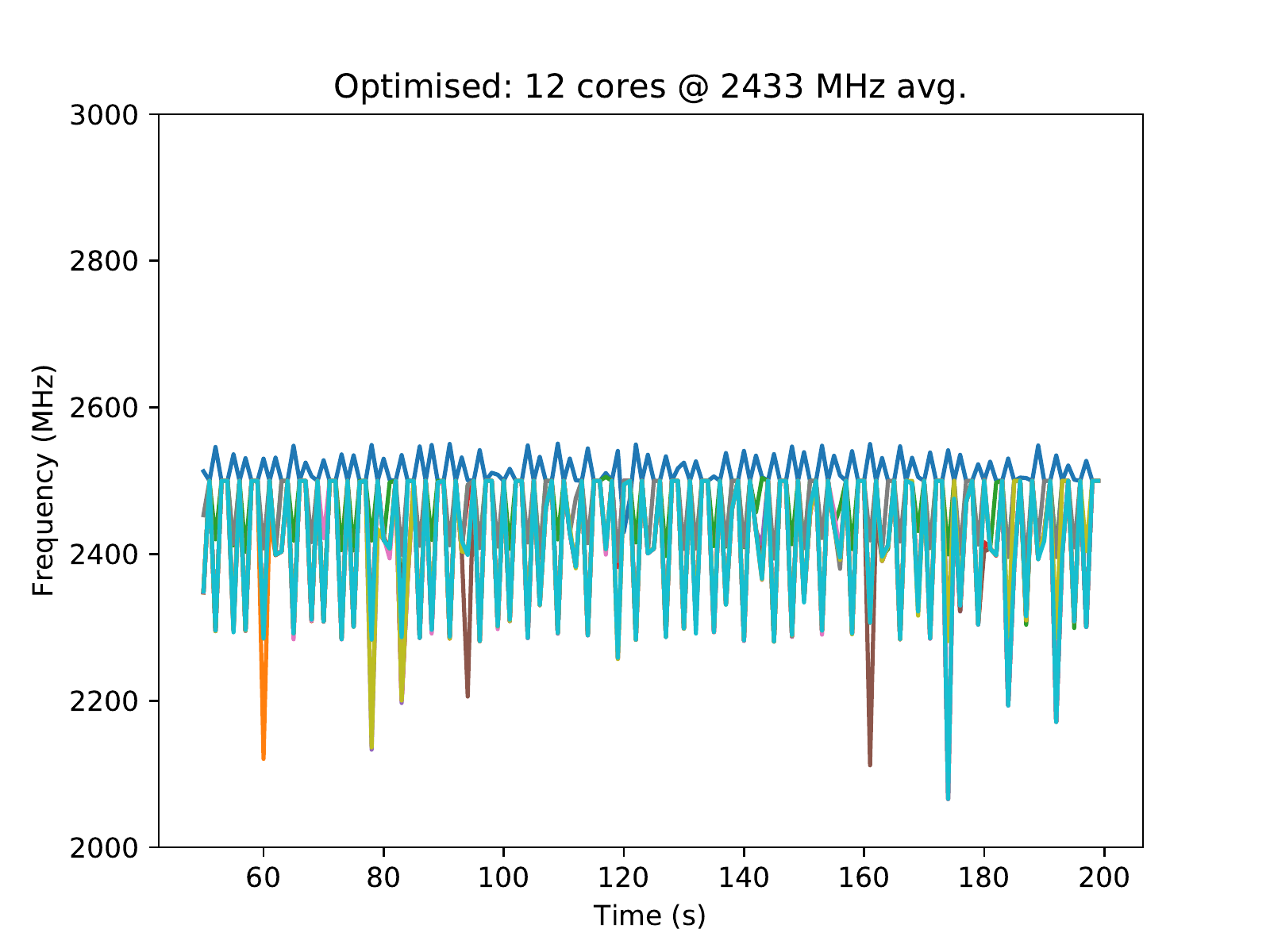}
  \includegraphics[width=0.3\textwidth]{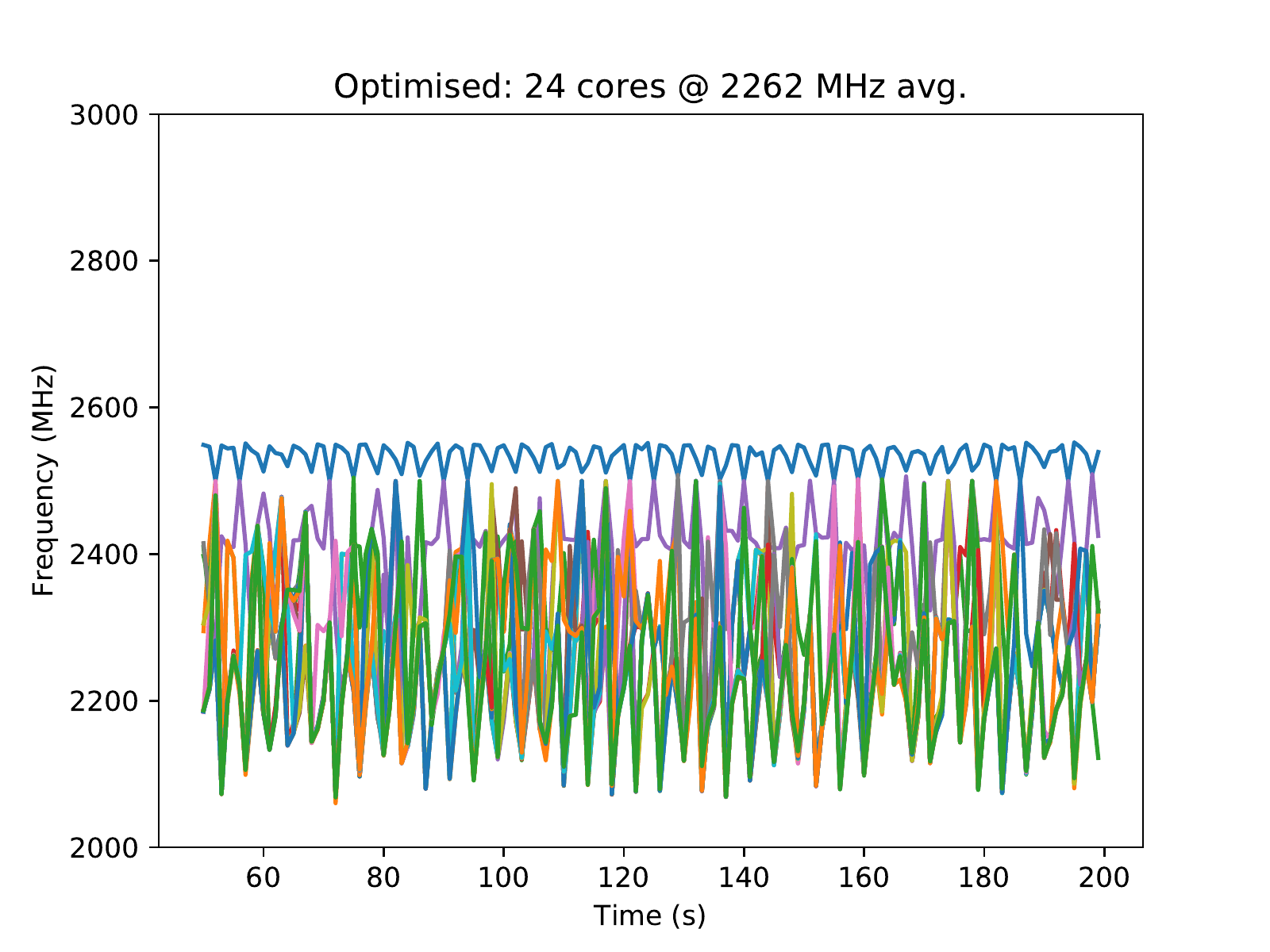}
 \end{center}
 \caption{
  Broadwell's 
  frequency choice per core for a dual core (left) a one-socket
  (middle) and a 24-core run (right).
  The caption gives the time-averaged frequency.
  All setups rely on code translated with AVX.
  Without the manual AVX optimisation, the average frequency is 2,616~GHz
  for two cores, drops to 2,454~GHz on a socket and finally to 2,409~GHz.  
  \label{figure:system:frequency}
 }
\end{figure*}

\textcolor{textR1}{
We kick off our experiments with studies on the Broadwell chip.
For statements on the code's scaling, it is important first to
understand the frequency behaviour under load.
On Broadwell, 
a single or dual core setup drives the chip at around 2.5~GHz
(Figure~\ref{figure:system:frequency}).
If we however use all 24 cores and run our optimised code variant using AVX,
the node is downclocked to around 2.165~GHz on average.
If we manually disable AVX, the downclocking is less severe (2.35~GHz on
average).
For all-core loads, overclocking is not used of the chip's own accord.
We observe one core---predominantly being busy with task production and
scheduling---to perform at close-to-nominal speed.
All others clock down.
Scalability graphs have to take the amortized downclocking into account.
}

\subsection{Scalability}
\label{subsection:measurements:scalability}

 \begin{figure}
  \begin{center}
   \centering
     \includegraphics[width=1.0\columnwidth]{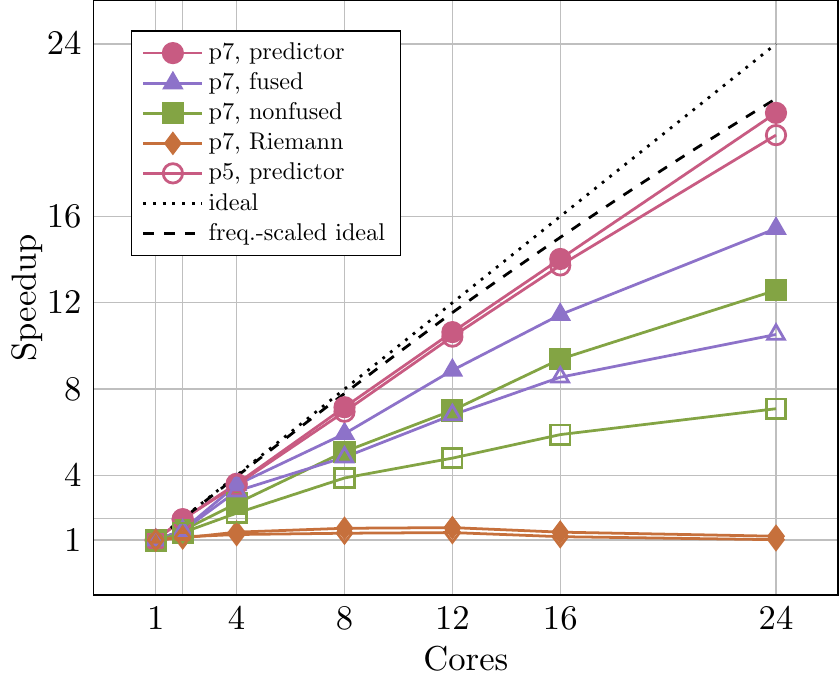}
\\
     \includegraphics[width=1.0\columnwidth]{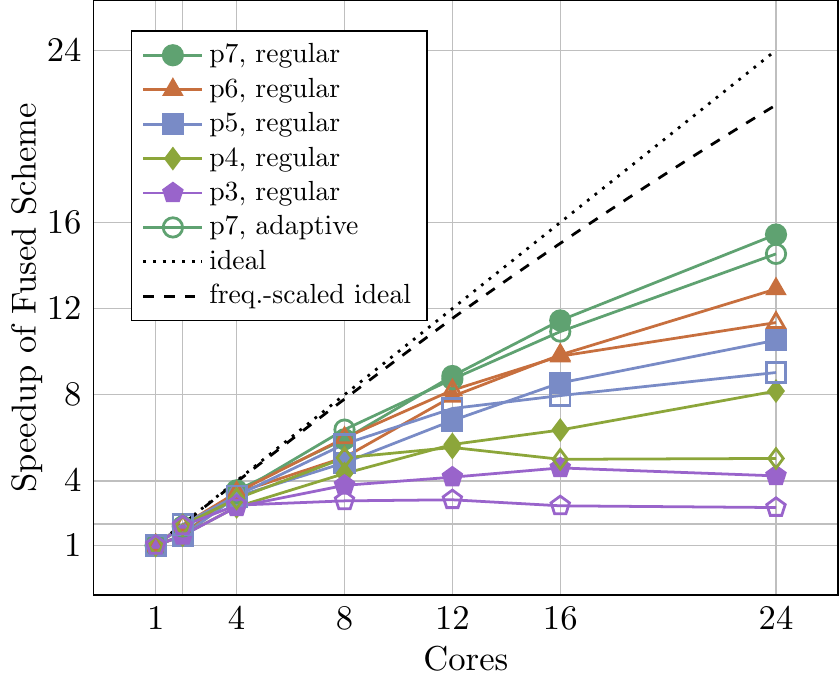}
  \end{center}
  \caption{
   Code scalability on Broadwell for various orders and a $27 \times 27 \times
   27$ grid.
   Top: 
   Scalability of the isolated predictor and Riemann phase
   of the nonfused scheme plus overall scalability of the nonfused
   (straightforward) and the fused schemes for orders 5 and 7.
   Both an ideal linear speedup and a speedup calibrated with the
   observed frequency reduction are given.
   Bottom: Scalability of the fused scheme 
   for orders 3---7. 
   Regular grid runs are compared to a adaptive
   grids where a dynamic refinement criterion is allowed
   to add additional grid levels to the regular base grid.
   \label{figure:code-measurements:scalability}
  }
 \end{figure}

\textcolor{textR1}{To assess the impact of frequency, horizontal 
\textcolor{textR2}{and} 
vertical diversity,} 
it would make limited sense to benchmark serial or non-scaling code.
Before we study our code's memory and energy characteristics, we
\textcolor{textR1}{thus} validate that the code scales reasonably
\textcolor{textR1}{on Broadwell}
(Figure~\ref{figure:code-measurements:scalability}).
\textcolor{textR1}{Qualitatively similar} results 
\textcolor{textR1}{arise on}
\textcolor{textR3}{the Intel Xeon Scalable chip}.
\textcolor{textR1}{
This holds despite the significantly changed memory architecture.
}

%
%
\textcolor{textR1}{
 As the 
}
corrector step is merged into the predictor in the fused scheme,
\textcolor{textR1}{
 we benchmark the fused scheme against the nonfused implementation and decompose
 the latter's behaviour into the scaling of the Riemann solve and the scaling of
 the cell-wise operator 
 \textcolor{textR2}{(Figure~\ref{figure:code-measurements:scalability}) top}.
 While the predictor scales perfectly for both $p=5$ and $p=7$ once we accept
 that the cores clock down, the Riemann solvers scale hardly at all.
 They are heavy on data movements, move many small chunks of data, and suffer
 from the AMR administration overhead which we did not remove from any plot.
 Furthermore, we exploit the first-touch policy to ensure that data is
 allocated following their cell associativity:
 all cell data plus all faces of this cell are allocated en bloc by the
 allocating thread.
 The Riemann solves however bring together data from two cells and thus suffer
 from NUMA effects.
 In the end, the combination of the three task types ends up in-between these
 two extreme cases.
 Fusion is robustly faster than the nonfused scheme.
}
%
%
The higher the order the more dominating the predictor
steps.
\textcolor{textR1}{
 The arithmetic intensity of the Riemann solves in contrast is close to
 $p$-invariant. 
}
We thus obtain better scaling overall
\textcolor{textR1}{
 when we increase the order.
}

\textcolor{textR1}{
Our experiments clarify that our code scales
reasonably on one socket.
The predictor ``saves'' the
overall scalability.
}
If we use the second half of cores too, the parallel efficiency
deteriorates.

\begin{observation}
 \textcolor{textR1}{
  Strongly dynamic AMR codes with heterogeneous tasks suffer from
  multi-socket architectures.
  It might be reasonable to deploy one process/rank per socket, i.e.~to give up
  on the idea of a larger shared memory system, \textcolor{textR2}{to reduce
  NUMA effects}.
 } 
\end{observation}

Higher orders mitigate this effect, while dynamic adaptivity makes it
\textcolor{textR1}{slightly} worse.
Dynamic adaptivity is not for free. 
Further increases of the mesh resolution $n$ (growing the
mesh with $\mathcal{O}(n^d)$ for regular grids)
or polynomial order (increase in $\mathcal{O}(n^p)$) are impossible due to
memory limits \textcolor{textR1}{on Broadwell}.
Our benchmark quickly \textcolor{textR1}{is caught in a}
strong scaling regime.
This observation is typical for many HPC codes working with dynamically adaptive
meshes.
They do not exhibit arbitrary concurrency.

\begin{observation}
 \textcolor{textR1}{
 To improve code scalability, increases of the
 polynomial order or mesh size are required. Such 
 increases however are constrained by the memory available.
 }
\end{observation}
 
\noindent
\textcolor{textR1}{
It does not come as a surprise that it is desirable to have more memory
to be able to increase either the resolution, i.e.~to shift the strong scaling
regime, or $p$, i.e.~to increase the arithmetic intensity
\citep{Hutchinson:2016:HighOrderSpecFEM}.
While this favors the introduction of novel large-scale memory as provided
through the Intel Optane technology, our observations suggest that any
architectural extension that increases NUMA penalties affects the overall performance negatively.
}

\subsection{Code characteristics and optimizations}
\label{subsection:measurements:code-characteristics}

%
%
If we rewrite our code into a fused variant where a task is immediately
triggered once its input data become available, we obtain
\textcolor{textR1}{faster} code.
It robustly pays off to issue compute tasks as soon as their input data is
available and thus to overlap computationally demanding with
memory-intense tasks.
This observation is in line with implicit data access blocking as we find it in
Intel's TBB \citep{Reinders:07:TBB}, where the task graph/tree is processed
depth-first.
\textcolor{textR1}{
 We thus focus solely on the fused scheme from hereon.
}

\begin{table*}[htb]
 \caption{
   Hardware counters on Broadwell (24 cores) for a $27^3$ grid.
   Scalar denotes no vectorization (\texttt{-no-vec -no-simd}) and no generation of vectorized inline assembler code.
   The top part presents the whole code characteristics if all three different
   task types are merged into each other.
   Below we split up the measurements into the phases prediction, Riemann solves
   and solution correction.
   \label{table:code-measurements:broadwell:hw-counters}
  }
  {\footnotesize
\begin{center}
\begin{tabular}{lrrrrrrrrrrrr}
\toprule
& \multicolumn{6}{c}{Regular} & \multicolumn{6}{c}{Dyn.~Adaptive} \\
\cmidrule(lr){2-7}
\cmidrule(lr){8-13}
& \multicolumn{3}{c}{Scalar} & \multicolumn{3}{c}{AVX2} & \multicolumn{3}{c}{Scalar} & \multicolumn{3}{c}{AVX2} \\
\cmidrule(lr){2-4}
\cmidrule(lr){5-7}
\cmidrule(lr){8-10}
\cmidrule(lr){11-13}
& \multicolumn{1}{c}{$p=3$} & \multicolumn{1}{c}{$p=5$} & \multicolumn{1}{c}{$p=7$} 
& \multicolumn{1}{c}{$p=3$} & \multicolumn{1}{c}{$p=5$} & \multicolumn{1}{c}{$p=7$} 
& \multicolumn{1}{c}{$p=3$} & \multicolumn{1}{c}{$p=5$} & \multicolumn{1}{c}{$p=7$} 
& \multicolumn{1}{c}{$p=3$} & \multicolumn{1}{c}{$p=5$} & \multicolumn{1}{c}{$p=7$} 
\\
\midrule
Avg. Time (s) &0.34&0.84&2.29&0.31&0.52&1.3&2.57&4.09&11.11&2.58&2.88&6.79\\
Gflop/s &12.3&32&41.7&12.8&47.9&74.3&3.7&18.4&37.1&5.7&23&58.1\\
L2 Request Rate (\%) &6.55&9.27&8.80&8.52&16.34&19.78&5.09&7.56&7.99&6.79&14.93&18.16\\
L2 Miss Ratio (\%) &13.76&14.43&18.95&10.37&15.28&19.96&17.12&16.91&18.26&11.92&13.63&17.59\\
L3 Request Rate (\%) &0.09&0.04&0.04&0.11&0.20&0.28&0.05&0.05&0.04&0.09&0.17&0.28\\
L3 Miss Ratio (\%) &29.38&7.43&8.85&60.69&20.91&13.40&28.23&14.15&9.86&44.55&24.26&11.57\\
Mem. Bandwidth (GB/s)  &7.9&6.5&5.4&6.4&12&9.7&4.5&6.8&5.7&6.1&8.2&9.4\\
\midrule  
Avg. Time (s) &0.3&0.51&1.78&0.28&0.38&0.94&1.95&2.56&9.04&1.97&2.32&4.59\\
Gflop/s &11.8&42&54.3&11.8&26.9&93.1&5.5&22.3&51.8&4&24.5&66.1\\
L2 Request Rate (\%) &6.49&8.69&8.92&9.21&17.17&18.17&5.08&8.43&8.63&7.14&14.24&18.32\\
L2 Miss Ratio (\%) &9.58&16.07&18.79&8.26&15.56&22.49&13.37&15.35&15.77&8.90&17.68&20.48\\
L3 Request Rate (\%) &0.02&0.04&0.04&0.05&0.19&0.28&0.03&0.04&0.03&0.06&0.16&0.25\\
L3 Miss Ratio (\%) &25.48&4.64&6.84&41.80&2.99&3.96&27.42&11.22&10.98&29.54&8.06&5.54\\
Mem. Bandwidth (GB/s)  &4.85&6.34&4.09&6.05&11.74&6.91&3.17&4.34&3.25&4.49&5.47&5.86\\
\midrule  
Avg. Time (s) &0.11&0.1&0.11&0.11&0.11&0.1&0.95&0.99&1.04&0.92&0.94&0.96\\
Gflop/s &0.6&3&9.4&0.6&3&8.6&0.5&3.8&15.6&0.2&1.1&2.8\\
L2 Request Rate (\%) &9.62&9.91&8.32&8.32&14.37&18.76&3.61&3.03&6.45&3.37&5.32&8.07\\
L2 Miss Ratio (\%) &17.18&16.42&17.15&16.65&16.07&19.71&20.43&20.08&5.29&23.54&20.26&21.85\\
L3 Request Rate (\%) &0.10&0.05&0.05&0.17&0.30&0.28&0.08&0.05&0.02&0.09&0.10&0.16\\
L3 Miss Ratio (\%) &30.97&21.59&12.72&33.23&18.92&9.96&24.76&23.86&19.27&28.34&26.17&14.61\\
Mem. Bandwidth (GB/s)  &2&2.9&4.3&1.66&2.4&3.31&3.7&3.32&2.71&3.36&3.51&3.62\\
\midrule 
Avg. Time (s) &0.11&0.14&0.18&0.14&0.14&0.18&1.13&1.55&2.1&1.1&1.27&1.72\\
Gflop/s &2&6.6&12.5&1.8&6.4&13&1.2&5.3&15.6&0.6&2.1&5\\
L2 Request Rate (\%) &5.61&6.51&8.10&16.06&19.20&25.79&3.69&3.53&7.22&5.59&9.01&13.64\\
L2 Miss Ratio (\%) &22.32&19.64&14.67&26.43&20.56&17.73&25.49&21.58&7.54&26.93&23.27&21.36\\
L3 Request Rate (\%) &0.10&0.05&0.05&0.32&0.34&0.42&0.08&0.08&0.04&0.13&0.24&0.26\\
L3 Miss Ratio (\%) &39.82&34.55&33.17&61.23&46.50&41.84&35.39&38.22&32.61&37.58&43.81&39.93\\
Mem. Bandwidth (GB/s)  &7.2&15.9&21.7&8.4&18&26.8&4.6&7.3&9.9&3.8&7.2&11.6\\

\bottomrule
\end{tabular}
\end{center}

  }
\end{table*}

%
%
Our benchmarking \textcolor{textR1}{continues} with performance counter
measurements.
Each test is done without any vectorization (disabled at compile time) 
and with full AVX2 vectorization.
We observe a robust speed improvement through vectorization (Table
\ref{table:code-measurements:broadwell:hw-counters}).
The reduction of the time-to-solution follows an increase of the Mflop
rate.
\textcolor{textR1}{
 The positive vectorization impact is solely due to the high order prediction
 tasks which make up for the majority of the runtime.
 The higher the polynomial order, the higher the fraction of the runtime plus
 the higher the Gflop/s. 
}

%
%
\textcolor{textR1}{
 While the predictors are responsible for the Gflop/s, the solution update,
 which is fused with the predictor, delivers the memory throughput. 
 It reaches around 25\% of Stream TRIAD \citep{McCalpin:95:Stream} on Broadwell. 
 Correcting the solution reduces the effective Mflop efficiency of this
 fused, cell-aligned task type. 
}
The vectorization \textcolor{textR1}{success} is diminished
\textcolor{textR1}{further} by the fact that
arithmetically intense tasks take turns with
\textcolor{textR1}{cheap Riemann solves}.
The \textcolor{textR1}{runtime of the} latter \textcolor{textR1}{does} not
benefit from vectorization \textcolor{textR1}{at all}.

\textcolor{textR1}{
In none of our experiments, we 
}
have been able to observe any significant impact of the task character
homogenization on the AVX \textcolor{textR1}{downclocking}.
One might guess that intermixing computationally heavy with cheap tasks implies
that not all cores run AVX at the same time, and the node thus does not throttle
the speed as significantly.
\textcolor{textR1}{
This however seems not to happen significantly. 
Figure \ref{figure:system:frequency} remains representative.
}

To \textcolor{textR1}{characterize} the cache usage, we measure per
cache the request and the miss rate
\textcolor{textR1}{(Table~\ref{table:code-measurements:broadwell:hw-counters})}.
Request rate means number of requests divided by number of instructions.
Miss rate means number of requests not served by a cache divided by number of
instructions.
From both rates, we can derive the miss ratio which is the ratio of cache
accesses which have not been served by a particular cache.

%
%
The request rate \textcolor{textR1}{of both L2 and L3} increases with increasing
polynomial order.
Furthermore, it is significantly higher for AVX-enabled code,
\textcolor{textR1}{while} 
the request rate decreases rapidly over the cache levels.
\textcolor{textR1}{
 Our code's aggressive cache blocking renders the dominating predictor
}
cache-efficient. 
\textcolor{textR1}{
 The predictor's data does not fit into L1, but barely any misses hit through
 the last-level cache (LLC).
}

%
%
Our \textcolor{textR1}{overall} miss ratio \textcolor{textR1}{however} is high:
Every time a piece of data is not found
in the \textcolor{textR1}{L2}, the \textcolor{textR1}{LLC} cannot serve this
request either \textcolor{textR1}{with high probability}.
\textcolor{textR1}{Though}
the miss ratio decreases with increasing polynomial order,
\textcolor{textR1}{
 high ratios imply that we are neither bandwidth- nor compute-bound.
 This problematic behaviour stems from the Riemann solves.
 They pollute the caches through their low arithmetic intensity, small input
 data cardinality and NUMA effects and thus both are cache-inefficient
 themselves and pollute the following correctors.
 It is the re-filling of the caches with small chunks of data for the
 corrector/predictor steps which slows down the code.
 It is dominated by memory latency. 
 The two steps themselves are memory-efficient.
}

\begin{table*}[htb]
 \caption{
   Measurements from Table~\ref{table:code-measurements:broadwell:hw-counters}
   for the Intel Xeon Scalable Gold running with all 36 cores. We show only data
   for the fused scheme without a breakdown into individual phases.
   \label{table:code-measurements:skylake:hw-counters}
  }
  {\footnotesize
\begin{center}
\begin{tabular}{lrrrrrrrrrrrr}
\toprule
& \multicolumn{6}{c}{Regular} & \multicolumn{6}{c}{Dyn.~Adaptive} \\
\cmidrule(lr){2-7}
\cmidrule(lr){8-13}
& \multicolumn{3}{c}{Scalar} & \multicolumn{3}{c}{AVX2} & \multicolumn{3}{c}{Scalar} & \multicolumn{3}{c}{AVX2} \\
\cmidrule(lr){2-4}
\cmidrule(lr){5-7}
\cmidrule(lr){8-10}
\cmidrule(lr){11-13}
& \multicolumn{1}{c}{$p=3$} & \multicolumn{1}{c}{$p=5$} & \multicolumn{1}{c}{$p=7$} 
& \multicolumn{1}{c}{$p=3$} & \multicolumn{1}{c}{$p=5$} & \multicolumn{1}{c}{$p=7$} 
& \multicolumn{1}{c}{$p=3$} & \multicolumn{1}{c}{$p=5$} & \multicolumn{1}{c}{$p=7$} 
& \multicolumn{1}{c}{$p=3$} & \multicolumn{1}{c}{$p=5$} & \multicolumn{1}{c}{$p=7$} 
\\
\midrule
Avg. Time (s) &0.28&0.38&1.15&0.27&0.32&0.95&2.32&3.01&6.48&2.52&3.19&5.36\\
Gflop/s &14.6&62.5&86.7&22&100.9&156.4&7.3&32&74.6&8&32.8&95.3\\
L2 Request Rate (\%) &4.82&5.55&7.64&9.31&15.64&21.68&5.81&5.66&8.76&11.06&18.78&22.00\\
L2 Miss Ratio (\%) &8.25&4.10&11.46&9.25&7.35&12.27&12.86&4.31&9.27&11.41&3.70&12.44\\
L3 Request Rate (\%) &0.03&0.02&0.03&0.11&0.06&0.16&0.06&0.01&0.03&0.15&0.08&0.17\\
L3 Miss Ratio (\%) &55.04&60.71&14.22&85.55&75.49&59.23&37.52&45.68&11.69&56.32&63.04&47.92\\
Mem. Bandwidth (GB/s)  &9.7&13.7&20.4&13.5&23.5&89.7&6.6&6.7&12.1&7.4&9.7&64.2\\

\bottomrule
\end{tabular}
\end{center}

  }
\end{table*}

\textcolor{textR1}{
 The Intel Xeon Scalable chip amplifies all observed trends.
 The chip delivers higher performance---also due to the increased core
 count---and benefits from a decreased L2 miss ratio
 (Table~\ref{table:code-measurements:skylake:hw-counters}), as
 the L2 cache per core is increased. 
 However, changing from inclusive to non-inclusive caches and reducing the
 cache-per-core size makes the code yield an even higher L3 miss ratio.
 This results in a significantly increased memory bandwidth.
}

\begin{observation}
\textcolor{textR1}{
The code suffers from memory latency.
}
\end{observation}

\noindent
\textcolor{textR1}{
 We obtain a reasonably high percentage of peak for a dynamically adaptive grid
 through the high order space-time predictor.
 The necessity to interwave it with cheap tasks however implies that we are
 overall neither compute- nor bandwidth-bound.
}

\section{Frequency and energy analysis}
\label{subsection:energy}

\begin{table*}
  \caption{
   Energy consumption per degree of freedom on Broadwell (24 cores) and Xeon
   Scalable (36 cores) for a typical run with a regular and a dynamic grid. 
   The total energy \textcolor{textR2}{plus the energy spent on the memory are
   given.}
   \label{table:code-measurements:energy}
  } 
\centering
{\footnotesize
\begin{tabular}{lrrrrrrrrrrrr}
\toprule
& \multicolumn{6}{c}{Regular} & \multicolumn{6}{c}{Dyn.~Adaptive} \\
\cmidrule(lr){2-7}
\cmidrule(lr){8-13}
& \multicolumn{3}{c}{Scalar} & \multicolumn{3}{c}{AVX2} & \multicolumn{3}{c}{Scalar} & \multicolumn{3}{c}{AVX2} \\
\cmidrule(lr){2-4}
\cmidrule(lr){5-7}
\cmidrule(lr){8-10}
\cmidrule(lr){11-13}
& \multicolumn{1}{c}{$p=3$} & \multicolumn{1}{c}{$p=5$} & \multicolumn{1}{c}{$p=7$} 
& \multicolumn{1}{c}{$p=3$} & \multicolumn{1}{c}{$p=5$} & \multicolumn{1}{c}{$p=7$} 
& \multicolumn{1}{c}{$p=3$} & \multicolumn{1}{c}{$p=5$} & \multicolumn{1}{c}{$p=7$} 
& \multicolumn{1}{c}{$p=3$} & \multicolumn{1}{c}{$p=5$} & \multicolumn{1}{c}{$p=7$} 
\\
\midrule
Total Energy (kJ) &1.46&5.24&15.54&1.29&3.02&9.01&27.57&41&96.84&22.24&28.42&53.7\\
Total Energy per DoF (mJ) &1.16&1.23&1.54&1.02&0.71&0.89&3.16&1.39&1.39&2.55&0.97&0.77\\
Energy DRAM (kJ) &0.49&1.25&3.06&0.43&0.78&1.79&6.96&9.03&19.12&5.42&6.62&10.64\\
Energy DRAM per DoF (mJ) &0.389&0.294&0.304&0.341&0.183&0.178&0.799&0.307&0.274&0.622&0.225&0.153\\

\midrule
Total Energy (kJ) &3.24&6.41&17.77&2.85&4.63&13.32&36.46&46.68&123.74&51.75&46.49&91.48\\
Total Energy per DoF (mJ) &2.57&1.51&1.76&2.26&1.09&1.32&4.19&1.59&1.78&5.94&1.58&1.31\\
Energy DRAM (kJ) &0.17&0.25&0.69&0.17&0.25&0.87&1.83&2.04&4.95&2.72&2.42&4.9\\
Energy DRAM per DoF (mJ) &0.135&0.059&0.068&0.135&0.059&0.086&0.210&0.069&0.071&0.312&0.082&0.070\\

\bottomrule
\end{tabular}

}
\end{table*}

AVX operations 
\textcolor{textR1}{
trade memory pressure for speed if data are accessed continuously,
}
while they in turn induce a frequency reduction if the chip exceeds its energy/temperature thresholds
\textcolor{textR1}{
(Figure~\ref{figure:system:frequency}).
}
Our results (Table~\ref{table:code-measurements:energy}) validate that
the reduction in time-to-solution compensates for
\textcolor{textR1}{problematic impact}:
With AVX, the executable delivers the results
faster at lower energy footprint.
This effect is the stronger the higher the polynomial order,
i.e.~the higher the arithmetic intensity of the heavy compute tasks.

\begin{figure}[htb]
  \begin{center}
   \centering
  \includegraphics[width=\columnwidth]{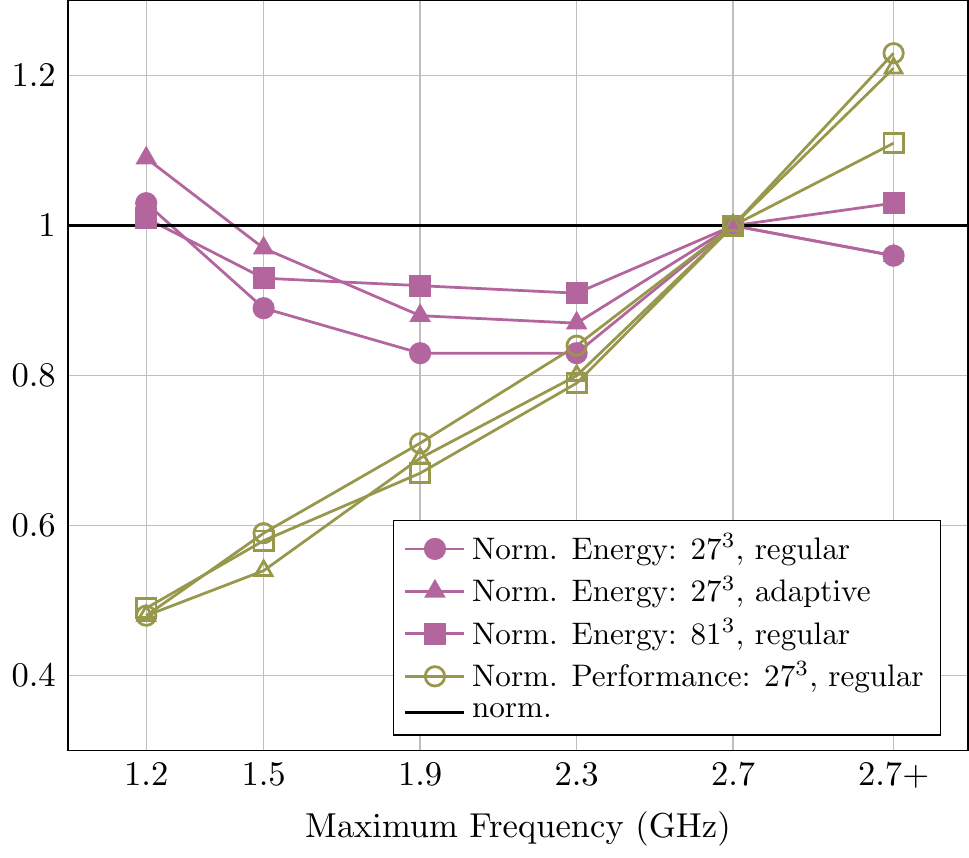}
  \end{center}
  \caption{
    Energy and time-to-solution results for various CPU frequencies on the
    Intel Xeon Scalable machine for $p=6$. 
    We normalize the results against the default frequency
    of 2.7~GHz. 
    $2.7+$ GHz denotes a base clock of 2.7~GHz with 3.7~GHz TurboBoost enabled.
    The memory frequency is determined by the chip (auto modus).
    \label{figure:code-measurements:energy}
   }
\end{figure}

We continue with experiments on our Xeon testbed and manually modify
the frequency of the chip.
Frequency alterations affect all standard components' maximum speed, while
intense AVX usage still might reduce the frequency.
Our data tracks the time-to-solution and the energy consumption per simulation
run (Figure~\ref{figure:code-measurements:energy}).

\begin{observation} 
 Running a chip at maximum frequency is best in terms of time-to-solution. If
 energy per simulation however is the optimality condition, a significant
 reduction of the frequency is advantageous.
\end{observation}

\noindent
Our results are in line with reports on the best-case efficiency for Linpack
if the total energy consumption has to be minimised \citep{Glesser:16:Frequency}.
\textcolor{textR2}{
They also agree with ADER-DG experiments on
tetrahedral meshes \cite{Breuer:15:EnergySeisSol}.
}
%
%
\textcolor{textR1}{
 More detailed studies however uncover three more insights:
 (i) A 
}
frequency alteration does not change the character of our
\textcolor{textR1}{latency} challenge.
\textcolor{textR1}{
 We observe no flattening of the speed curve
 when we reduce the frequency.
 Notably, we have not
 \textcolor{textR2}{been able} 
 to observe any statistically significant
 impact
 \textcolor{textR2}{on the cache counters and thus latency effects when we did
 alter} the memory speed against the CPU (not shown). 
 The memory's auto mode choosing an appropriate memory speed is not outshadowed
 by any manual memory frequency tuning.
 (ii)
 For our heterogeneous task graphs, turbo boost techniques which allow cores to
 temporarily upclock yield significantly improved performance at a limited
 increase of energy hunger.
 (iii)
 With cache capacity moving closer to the chip, our cache-optimized algorithm
 also makes the chip spend more energy on the cores rather than the memory
 (Table \ref{table:code-measurements:energy}).
}

\begin{observation} 
 \textcolor{textR1}{
 Core frequency reductions---whether manually imposed or triggered through
 AVX---are \emph{not} sufficient to mitigate latency effects.
 } 
\end{observation}

\section{Pinning, hyperthreading and latency hiding}
\label{subsection:pinning}

\textcolor{textR1}{Many HPC codes report pinning to be}
essential to achieve reasonable performance
\textcolor{textR1}{and to avoid NUMA pollution}.
We have not been able to confirm that pinning \textcolor{textR1}{pays off} 
for our code.
No data are presented here, as no statistically pinning impact could be
observed:

\begin{observation}
Runtimes with and without pinning can hardly be distinguished.
\end{observation}
 
\noindent
As our code is extremely cache efficient, data resides in the cache close to the
core.
If the system should decide to migrate a running task, the cache content has to
be moved, too.
However, a code with such extremely localized data access usually does not run
into traditional cache conflicts and false sharing.

Hyperthreading and oversubscription are a popular technique for deep
memory hierarchies (Figure~\ref{figure:introduction:optane-imdt}) 
\textcolor{textR1}{or} systems
where floating point units are shared between physical threads:
Data requests by a thread which cannot served by a near cache trigger cache
transfers.
Meanwhile, the system swaps this thread with another compute thread until
the data eventually has arrived.
\textcolor{textR1}{
 The compute facilities thus do not idle.
}
This is similar to the streaming/\textcolor{textR1}{high throughput}
 compute paradigm in CUDA.
While one thread uses the floating point capabilities, other threads can
fetch/prepare all data for the subsequent AVX usage and queue to continue once
the first thread ``releases'' the vector units.

\begin{figure}
 \begin{center} 
 \includegraphics[width=\columnwidth]{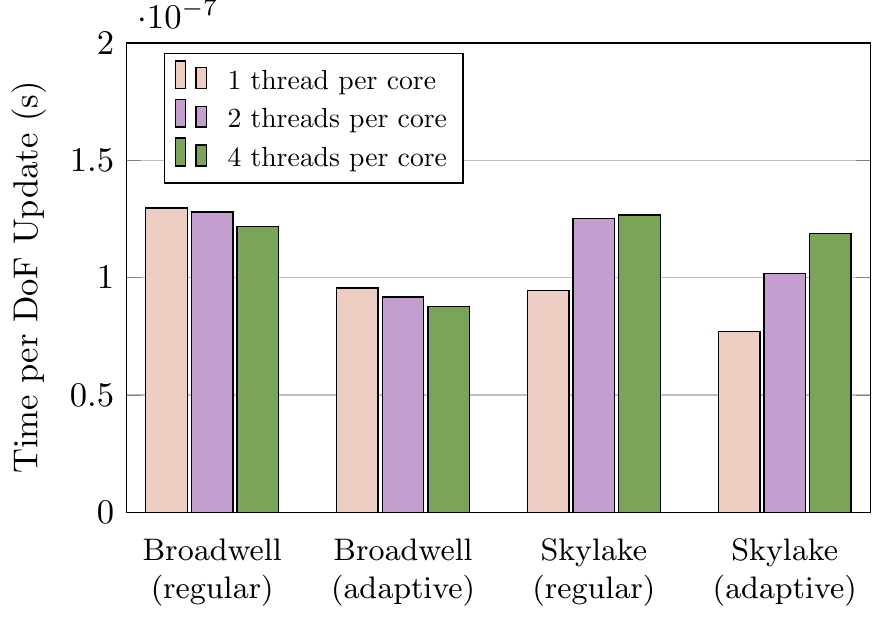}
 \end{center}
 \caption{
 \label{figure:oversubscription}
 Characteristic $p=7$ runs for both regular and adaptive meshes on the 36 cores
 of the Intel Xeon Scalable.
 We benchmark a one thread per core setup to oversubscribing with 2
 (hyperthreading) or even 4 threads.
 The latter overbooks each hyperthread with two logical threads.
 The label Skylake identifies the Intel Xeon Scalable processor.
 }
\end{figure}

\textcolor{textR1}{
On Broadwell,
}
the techniques do yield performance improvements in our case
(Figure~\ref{figure:oversubscription}).
\textcolor{textR1}{
 Starting from the Intel Xeon Scalable (Skylake) hardware generation however,
 they are counterproductive.
}
They decrease the performance.
This holds for all problem sizes.

\begin{observation}
 Our code's performance suffers from hyperthreading and thread oversubscription
 \textcolor{textR1}{
 on systems with non-inclusive caches.
 Oversubscription is \emph{not} a way forward to mitigate latency effects here.
 }
\end{observation}

\noindent
The result is not a surprise
\textcolor{textR1}{once we take} 
into account that the code relies heavily on
data access localization and
\textcolor{textR1}{
 that individual tasks with their high arithmetic intensity already fill the
 close caches.
}
Within one task, the code streams data through the AVX components.
\textcolor{textR1}{
 Switching tasks is
}
expensive:
\textcolor{textR1}{
 It interrupts
}
the AVX usage pattern \textcolor{textR1}{of} the current thread
\textcolor{textR1}{
 and induces further capacity misses on the close-by caches.
 If the data then still resides in a next-level cache, these runtime penalties
 are eventually compensated by the gain in vector facility utilization.
 If swapped-out data however is not contained in a close-by cache---a situation
 likely with non-inclusive caching---swapping out logical threads becomes too
 expensive to be compensated. 
}
\textcolor{textR2}{
 The effect is amplified by overhead necessary to administer the task queues and
 sequentialization and synchronization effects stemming from work stealing, e.g.
}

%

\section{Additional deep memory (Intel Optane technology)}
\label{subsection:optane}

%
%
Finally, we scale up our problem size such that it does not fit into 
our conventional memory anymore.
We use \textcolor{textR3}{Intel Optane technology} to accommodate this larger
memory footprint.
 Hence, the conventional memory becomes a fully associative L4
 cache. 
 It is Intel's hardware that is responsible for bringing data into and
 out of the cache.
 Agnostic of the particular data movement strategy, we may assume that high
 temporal and spatial locality \citep{Kowarschik:03:CacheTechniquesOverview} in
 the memory accesses continues to be advantageous.
 Agnostic of the specific hardware properties, we may assume that 
 ``main memory cache misses'' suffer from \textcolor{textR3}{higher}
 latency.

 
%
%
 
\begin{figure}
  \begin{center}
 \includegraphics[width=\columnwidth]{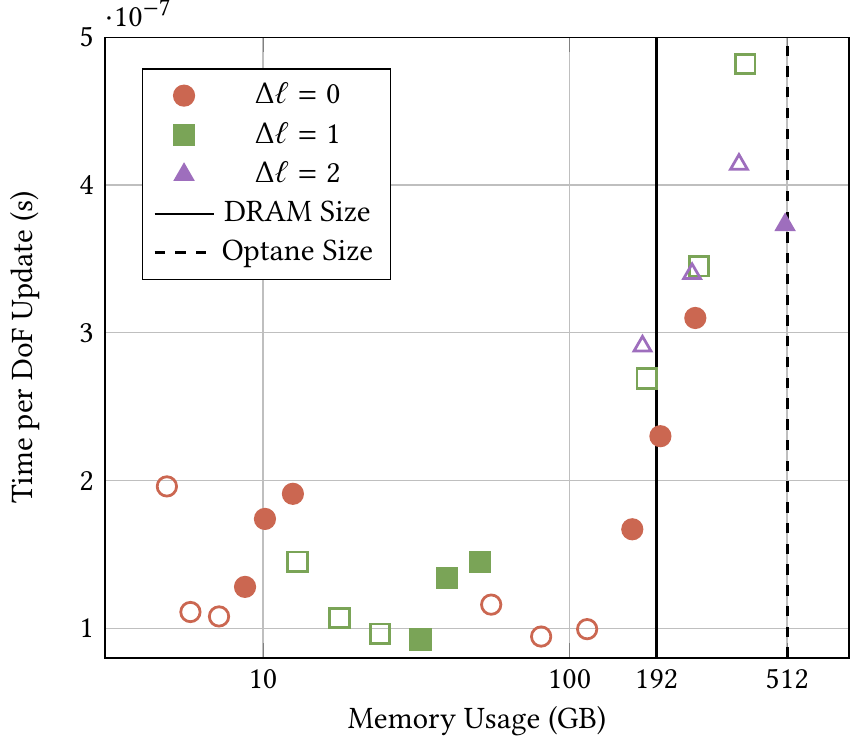}
 \includegraphics[width=\columnwidth]{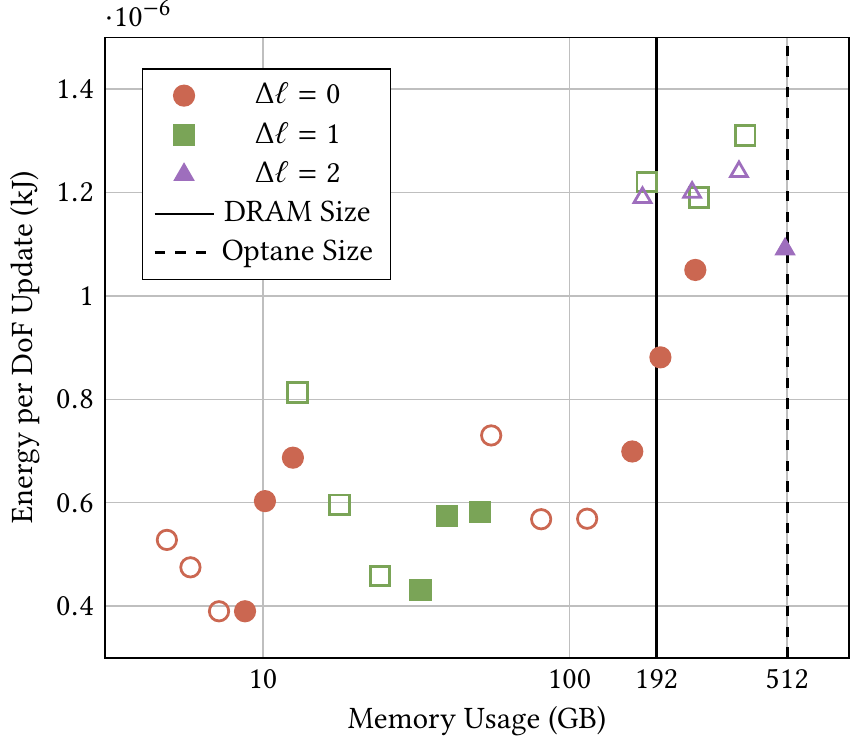}
  \end{center}
  \caption{
   Intel Xeon Scalable experiments for various problem sizes.
   Setups left of the dotted line use no Intel Optane technology as everything
   fits into the DRAM, so we switch if off.
   Top: Time per degree of freedom update. 
   Bottom: Total energy usage. 
   Each setup is, as long as it fits into the memory, computed 
   \textcolor{textR2}{multiple times (from left to right):
     We start from a base grid of $27 \times 27 \times 27$ (circles) and test 
     six
     polynomial orders $p \in \{4,5,6,7,8,9\}$ with $p \leq 6$ denoted through
     empty symbols.
     We then add one level of adaptivity ($\Delta \ell =1$, squares).
     We next rerun the regular grid experiment with a base grid of $81 \times
     81 \times 81$ (circles) and one level of adaptivity (squares).
     A base grid of $27 \times 27 \times 27$ with two levels of adaptivity
     (triangles) already requires Intel Optane technology unless we choose
     $p=3$.
   }
   \label{figure:code-measurements:optane}
  }
\end{figure}

%
%
For the benchmarking, we 
start from two regular grids: $27^3$ and $81^3$.
These regular grids are denoted by $\Delta \ell =0$, as we add 0 levels of
dynamic adaptivity.
\textcolor{textR1}{
 Different to previous setups where, for reasonably large setups, memory
 constrains the dynamic adaptivity criterion and allows it to add at most one level of grid
 refinement,} \textcolor{textR2}{we now} also conduct experiments with up to
 two additional resolution levels of \textcolor{textR1}{dynamical adaptive}
 meshes \textcolor{textR1}{($\Delta \ell \in \{1,2\}$)}.
All data is \textcolor{textR1}{normalized} against the real degrees of freedom
\textcolor{textR1}{
 and the number of time steps, as explicit hyperbolic equation solvers require
 the time step size to scale with the (adaptive) mesh size.
 We measure the cost per degree of freedom (DoF) updates per time step.
 As we found the large setups to crash with multicore support---the per-thread
 call stack was exceeded---we made each thread outsource its significant
 temporary local data structures to the heap.
 This is less efficient that on-stack storage and reduces the scalability,
 yet has to be done for all experiments here to obtain consistent data.
 Future versions of TBB will fix this stack size problem.
}

 
We \textcolor{textR1}{spot} a v-shaped cost profile for the setups fitting
completely into memory (Figure~\ref{figure:code-measurements:optane}).
The cost per DoF update increases with $p$, an effect in practice more than
compensated through the higher order of the approximation.
\textcolor{textR2}{
 Further to that, we see that this increase
 is more than made up as long as $p\leq 6$.
}
Vector
units are used more efficiently (cf.~Tables~\ref{table:code-measurements:broadwell:hw-counters}
and \ref{table:code-measurements:skylake:hw-counters}).
Once $p \geq 7$, the increase in cost also materialises in increased
runtimes.
\textcolor{textR2}{
 The memory of the individual compute steps exceeds close caches
 (cf.~L2 Request Rate in \ref{table:code-measurements:skylake:hw-counters}).
 We start to suffer from L2 or L3 cache misses. 
}
The v-pattern translates into energy per DoF update, too.

%
%
Once our regular grid setup exceeds main memory, 
we 
\textcolor{textR2}{experience}
a runtime penalty.
\textcolor{textR2}{
 For the low order experiments, this penalty is significantly below a factor of
 three which would mirror }
\textcolor{textR3}{the fact that the hardware has higher latency, too.}
\textcolor{textR2}{
 With increasing orders, the penalty increases.
}

\begin{observation}
  Trading bandwidth for latency does \emph{not} work for our code. 
  We suffer directly from increased latency.
\end{observation}

%
%
\noindent
\textcolor{textR2}{
 With the Intel Optane technology,
}
the v-pattern is distorted.
The higher our bandwidth demands, the higher also the LLC misses and
the higher the runtime penalty of the Intel Optane technology.
We observe that the most aggressive adaptivity pattern $\Delta \ell =2$ 
now yields a better cost per DoF update ratio than the more regular
discretizations.
The higher the polynomial order the more significant this effect.
As the dynamically adaptive mesh intermixes memory-intense and
arithmetically demanding tasks stronger, the runtime penalty induced by the
Intel Optane technology is more
significant for the other grid setups.
\textcolor{textR2}{
 For $\Delta \ell=1$, we have not been able to reproduce this effect which might
 be due to the fact that we ran into memory limits.
}

\begin{observation}
  We find the simulation for dynamically adaptive meshes being better suited to
  Intel Optane technology than a regular grid/fixed mesh setup.
\end{observation}

\noindent
We consider it to be a pattern of many important HPC codes that
data access exhibits stream access behaviour close to the compute core---here
notably for high orders.
On a higher abstraction level, codes however rely on flexible, dynamic tasks and
thus do not fit to hardware tailored to stream access.
Yet, with many cache levels, the arising non-local data accesses do not hammer
the last level memory that often.
As long as not too many codes access the memory concurrently, the latency
penalty remains under control.
Runtimes should thus intermix computationally heavy and memory-demanding tasks.

While our code exhibits no clear correlation of adaptivity pattern and
polynomial order to energy cost in the main memory, we do observe that the usage
of Intel Optane technology increases the energy footprint.
Future work will have to analyze whether persistent memory modes can bring down
these increased energy cost.

\section{Summary and conclusion}
\label{section:conclusion}


\textcolor{textR1}{
 Our manuscript studies a non-trivial solver for partial differential equations
 as we consider it to be characteristic for many upcoming simulation codes:
 it relies on many tasks of different compute character;
 the runtime of the
 tasks \textcolor{textR2}{and} the task composition are hard to
 predict---adaptive mesh refinement \textcolor{textR2}{
 plays a major
 role here and the situation might become more severe once non-linear equations
 are solved which require localized Newton or Picard iterations;
 finally,} and the efficiency of the solve hinges on the opportunity to use high polynomial orders and fine meshes.
 The solver requires massive memory.
}

Computer memory designers operate in a magic triangle of size, bandwidth and latency.
Under given energy and cost constraints, not all three of these
characteristics can be improved.
While caches optimize for bandwidth and latency, the new Intel memory
optimizes for size and bandwidth,
\textcolor{textR1}{
 while a core increase amplifies NUMA effects---notably for low-order and cheap
 (Riemann) tasks.
 As we find our code suffer from memory latency in general, we
 hypothesized that it may pay off to reduce the core frequency relative to the memory
 frequency, to use core oversubscription, to hide latency penalties, and to
 regularize and homogenize all computations, i.e.~to work with as regular data
 structures and task graphs as possible.
 We have not been able to confirm these hypotheses
}
\textcolor{textR2}{
 in general.
 However, we have found or confirmed attractive alternative solutions or
 solution proposals per diversity axis.
}

%

\textcolor{textR1}{
 An increasing flexibility and heterogeneity of clock frequencies allows chips
 to alter the frequencies for individual system parts.
 We have not been able to exploit this feature to soothe the impact of latency, 
 although we have confirmed the well-known insight that drastically decreased
 frequencies improve the energy efficiency of the simulation.
 Yet, our data suggests that the turbo boost feature of modern chips is of use
 for very heterogeneous task graphs. 
 It significantly improves the runtime while the energy demands remain under
 control.
 \textcolor{textR2}{
 It}
 might be reasonable to downclock chips overall, but to allow
 the runtime to increase the frequency of particular cores starting from the
 reduced baseline up to the turbo boost frequency.
 Notably those cores producing further tasks and running computationally cheap
 tasks would benefit from such a feature.
 \textcolor{textR2}{
  Such a fine-granular frequency alteration feature---likely coupled with a task
  runtime---seems to be promising.
  }
}

\textcolor{textR1}{
 An increasing core count and thus NUMA heterogeneity amplify NUMA effects
 which we label as growing horizontal diversity.
 Our data suggests that it might be reasonable to subdivide large shared memory
 chips into logically distributed memory systems.
 We propose to place multiple MPI ranks on each node.
}
Our benchmarks \textcolor{textR1}{furthermore} clarify that existing cache
optimization techniques---notably a high data access localization---\textcolor{textR1}{continue to} pay off.
They help to soothe the impact of massively increased latency.
\textcolor{textR1}{
 In return, however, overbooking is not an option to hide
 latency/NUMA effects as vendors give up on inclusive caching.
 It is future work to study whether runtimes explicitly copying data from
 persistent/large-scale memory into the ``right'' part of the main
 memory---which effectively becomes the LLC---can help to eliminate NUMA effects
 \citep{Kudryavtsev:17:NUMAOptane}.
 In this case, future runtimes have to be equipped with the opportunity to
 predict the task execution pattern and to replace classic prefetching with
 explicit memory moves.
}

In the case of ExaHyPE, \textcolor{textR1}{a homogenized task}
parallelism \textcolor{textR1}{which mixes} tasks of different
compute characteristics allowed us 
to \textcolor{textR1}{hide} some latency of the Intel Optane technology.
\textcolor{textR1}{
 Our code has been able to cope with the increased vertical memory diversity. 
} 
We show that it is absolutely essential to equip tasking systems and algorithms
with the opportunity to run memory-intense and compute-bound tasks concurrently,
while the majority of compute-intense jobs has to exhibit data access locality.
If we get the balance between bandwidth and compute demands right, latency
effects remain under control.
The access pattern has to be homogenized.
Future task systems should internally be sensitive to the compute
character of the tasks. They have to mix compute-intense jobs with memory-intense jobs to
avoid that a whole node waits for slow deep memory. This naturally can be
mapped onto job priorities and mechanisms ensuring that not too many jobs of
one priority are launched.
\textcolor{textR1}{To the best of our knowledge, current runtimes as found with
OpenMP, TBB or C++11 lack}
mature support for \textcolor{textR1}{such} priorities \textcolor{textR1}{or
constraints}.

Machines equipped with Intel Optane technology provide ample memory. 
It is an appealing alternative to classic ``fat nodes''; also
in terms of procurement cost.
Once the exascale era makes the total power budget of computers grow to tens of
megawatts, it is an option to trade, to some degree, the DRAM for an 
energy-modest extra layer of memory.
\textcolor{textR2}{
 This paper's experiments navigate at the edge of ``fits into the memory'' and
 thus provides too few experimental samples to support claims through frequently
 observed patterns.
 We need to run more experiments with more hardware configurations and more
 applications. 
 Yet, our results
}
\textcolor{textR1}{
 suggest that the way forward
\textcolor{textR2}{
 into the massive-memory age
}
  might not be a na\"ive rendering of
 the main memory into an additional cache layer;
 at least not for non-trivial/non-streaming codes.
}
\textcolor{textR2}{
 Instead, we ask for three architectural or software extensions: fine-granular
 frequency control, runtimes with explicit data prefetching, and runtimes with
 mature task priorities; 
 the latter perhaps even guided by the availability of task data in close
 caches.
}


\begin{acks}
 The authors appreciate support received from the European Union Horizon 2020
 research and innovation programme under grant agreement No 671698 (ExaHyPE).
 This work made use of the facilities of the Hamilton HPC Service of Durham
 University.
 Particular thanks are due to Henk Slim for supporting us with Hamilton.
 Thanks are due to all members of the ExaHyPE consortium who made this research
 possible;
 notably and J.-M.~Gallard for integrating aggressively optimised
 compute kernels into ExaHyPE, and K.~Duru, A.-A.~Gabriel \textcolor{textR2}{as
 well as} L.~Rannabauer for realising a seismic benchmark on top of ExaHyPE.
 The authors are particular thankful to Leonhard Rannabauer for the support on
 running the seismic benchmarks.
 All underlying software is open source \citep{Software:ExaHyPE}.
\end{acks}

\bibliographystyle{SageH}
\bibliography{paper}

\end{document}